\documentclass[11pt]{article}
\usepackage{amssymb}
\usepackage{graphics}
\usepackage{epsfig}
\usepackage{a4wide}
\begin{document}
\title {Full ${\cal O}(\alpha_{ew})$ electroweak corrections to
$e^+e^- \rightarrow H H Z$
\footnote{Supported by National Natural Science Foundation of China.}} \vspace{3mm}
\author{{ Zhang Ren-You$^{2}$, Ma Wen-Gan$^{1,2}$, Chen Hui$^{2}$,
Sun Yan-Bin$^{2}$, and Hou Hong-Sheng$^{2}$}\\
{\small $^{1}$ CCAST (World Laboratory), P.O.Box 8730, Beijing
100080, P.R.China}\\
{\small $^{2}$ Department of Modern Physics, University of Science and Technology}\\
{\small of China (USTC), Hefei, Anhui 230027, P.R.China}}
\date{}
\maketitle \vskip 12mm

\begin{abstract}
We calculate the full ${\cal O}(\alpha_{ew})$ electroweak
corrections to the Higgs pair production process $e^+ e^-
\rightarrow H H Z$ at an electron-positron linear collider in the
standard model, and analyze the dependence of the Born cross
section and the corrected cross section on the Higgs boson mass $m_H$
and the c.m. energy $\sqrt{s}$.
To see the origin of some of the large corrections clearly, we calculate
the QED and genuine weak corrections separately.
The numerical results show that
the corrections significantly suppress or enhance the Born cross
section, depending on the values of $m_H$ and $\sqrt{s}$.
For the c.m. energy $\sqrt{s} = 500~ {\rm GeV}$, which is the
most favorable colliding energy for $H H Z$ production with
intermediate Higgs boson mass, the relative correction decreases
from $-5.3\%$ to $-11.5\%$ as $m_H$ increases from $100$ to
$150~ {\rm GeV}$. For the range of the c.m. energy where the cross
section is relatively large, the genuine weak relative correction
is small, less than $5\%$.

\end{abstract}

\vskip 5cm
{\large\bf PACS: 12.15.Lk, 14.80.Bn, 14.70.Hp, 11.80.Fv} \\
{\large\bf Keywords: electroweak correction, Higgs self-coupling,
Higgs pair production}

\vfill \eject \baselineskip=0.5cm      
\renewcommand{\theequation}{\arabic{section}.\arabic{equation}}
\renewcommand{\thesection}{\Roman{section}}
\newcommand{\nb}{\nonumber}
\makeatletter      
\@addtoreset{equation}{section}
\makeatother       

\section{Introduction}

\par
One of the most important goals of present and future colliders is
to study the electroweak symmetry breaking mechanism and the
origin of the masses of massive gauge bosons and fermions. As we
know, within the Higgs mechanism \cite{Higgs mechanism} the
electroweak gauge fields and fundamental matter fields (quarks and
leptons) acquire their masses through the interaction with a
scalar field (Higgs field) which has an nonzero vacuum expectation
value (VEV). And the self-interaction of the Higgs field induces
the spontaneous breaking of the electroweak ${\rm SU}(2)_{L}
\otimes {\rm U}(1)_{Y}$ symmetry down to the electromegnetic ${\rm
U}(1)_{EM}$ symmetry.

\par
The present precise experimental data have shown an excellent
agreement with the predictions of the standard model(SM) except
for the Higgs sector \cite{SM experiments}. These data
strongly constrain the couplings of the gauge boson to fermions
($\lambda_{Z f \bar{f}}$ and $\lambda_{W f \bar{f}^{\prime}}$),
and the gauge self-couplings, but say little about the couplings of
the Higgs boson to fermions ($\lambda_{H f \bar{f}}$) and gauge
bosons ($\lambda_{H Z Z}$ and $\lambda_{H W W}$). Recent LEP2
experiment suggests that the Higgs boson has the mass with a lower
bound of $114.4~ {\rm GeV}$ and a upper bound of $193~ {\rm GeV}$
at the $95\%$ confidence level \cite{lower bound on mH} \cite{up bound
on mH}. People believe that with the help of future high energy
colliders, such as the CERN Large Hadron Collider (LHC) and Linear
Colliders (LC's), precise tests of the Higgs sector become
possible. In the past few years, many theoretical works have been
contributed to studying the Higgs boson productions and the
properties of Higgs couplings at future high energy colliders
\cite{Higgs production LC} \cite{Higgs production LHC}.

\par
To reconstruct the Higgs potential and verify the Higgs mechanism
experimentally, not only the Yukawa couplings and the couplings of
the Higgs boson to gauge bosons should be measured, but also the
Higgs self-couplings which include the trilinear coupling
$\lambda_{HHH}$ and the quartic coupling $\lambda_{HHHH}$. These
Higgs self-couplings can be probed directly only by multi-Higgs
boson production. Due to the fact that the cross sections for
three Higgs boson production processes are much smaller than those
for Higgs boson pair production \cite{smaller-LC-1}
\cite{smaller-LC-2}, the quartic Higgs self-coupling remains
elusive. Recently, the Higgs boson pair production processes have
been widely considered, and the cross sections for these processes
in the SM have been evaluated at linear colliders and hadron
colliders. The phenomenology calculations show that it would be
extremely difficult to measure the Higgs self-coupling
$\lambda_{HHH}$ at the LHC \cite{Baur},
and $e^+ e^-$ linear colliders, where the study of the
$e^+ e^- \rightarrow H H Z$ and $H H \nu \bar{\nu}$ can be performed
with good accuracy, represent a possibly unique opportunity for
performing the study of the trilinear Higgs self-coupling \cite{smaller-LC-1}.
For the center of mass (c.m.) energy $\sqrt{s}$
from $500~ {\rm GeV}$ up to about 1 TeV, the $H H Z$ production with
intermediate Higgs boson mass is the most promising process among the
various Higgs double-production processes, since its cross section
is relatively large and all the final states can be identified
without large missing momentum. When the c.m. energy $\sqrt{s}$
exceeds 1 TeV, the $e^+ e^- \rightarrow H H \nu \bar{\nu}$ process
becomes sizeable, and it is possible to measure the trilinear
Higgs self-coupling $\lambda_{HHH}$ by using this process.
Therefore, in the first stage of a LC ($\sqrt{s} < 1~ {\rm TeV}$),
$e^+ e^- \rightarrow H H Z$
is the most promising channel to measure the Higgs self-coupling
$\lambda_{H H H}$.

\par
Although the cross section for $e^+ e^- \rightarrow H H Z$ with
intermediate Higgs boson mass is only about $0.1 \sim 0.2$ fb for
$\sqrt{s} < 1~ {\rm TeV}$, the measurement of the Higgs
self-coupling $\lambda_{H H H}$ through the process at $e^+ e^-$
colliders can be significantly improved. For example, C. Castanier
et al, conclude that a precision of about $10 \%$ on the total
cross section for $e^+ e^- \rightarrow H H Z$ can be achieved
leading to a relative error on $\lambda_{HHH}$ of $18\%$ with the
help of high integrated luminosity $\int {\cal L} = 2~ {\rm
ab}^{-1}$ after performing the detailed simulations of signal and
background processes at the TESLA\cite{simu-1}. Other simulations
demonstrate that the Higgs self-coupling $\lambda_{HHH}$ can be
extracted more accurately by using a discriminating variable,
namely the invariant mass of the HH system, and one can expect the
high sensitivity of the triple Higgs self-coupling with an accuracy to
$8\%$ and better in multi-TeV $e^+ e^-$ collisions\cite{simu-2}.
Therefore, to determine the Higgs self-couplings and reconstruct
the Higgs potential, the theoretical prediction of the cross
sections for $e^+ e^- \rightarrow H H Z$ at a LC within per-cent
accuracy is necessary. For this purpose, we investigate the $e^+
e^- \rightarrow H H Z$ process at a LC in detail and present the
calculation of the cross section for $e^+ e^- \rightarrow H H Z$
with the full ${\cal O}(\alpha_{ew})$ electroweak corrections in
the SM in this paper.

\par
The paper is organized as follows: in section \ref{sec-calculations} we
present the calculations of the Born cross section for $e^+ e^- \rightarrow H H Z$
and the full ${\cal O}(\alpha_{ew})$ electroweak corrections to this process.
The numerical results and discussions are presented in section \ref{sec-results}.
In the last section, a short summary is given.

\section{Calculations}
\label{sec-calculations} In this paper we adopt the 't
Hooft-Feynman gauge of the SM. At the tree level, there are six
Feynman diagrams relevant to the process $e^+ e^- \rightarrow H H
Z$ ( shown in Fig.1). In Fig.1 only the second Feynman diagram
(Fig.1(b)) contains a trilinear Higgs self-coupling vertex. In the
SM the Higgs potential can be expressed as
\begin{eqnarray}
V = \frac{m_H^2}{2} H^2 + \frac{m_H^2}{2 v} H^3 + \frac{m_H^2}{8 v^2} H^4
\end{eqnarray}
where $v = \left(\sqrt{2} G_F\right)^{-1/2}$. The trilinear Higgs
self-coupling constant, $\lambda_{HHH}$, can be derived from this
potential directly. By using the Higgs self-coupling constant
$\lambda_{HHH} = 3 m_H^2/v$ and the relevant Feynman rules for
gauge interactions, we can obtain the tree level amplitude ${\cal
M}_{{\rm tree}}$ and the cross section $\sigma_{{\rm tree}}$ for
$e^+ e^- \rightarrow H H Z$.

\begin{figure}[htbp]
\centering
\scalebox{0.75}{\includegraphics*[155,347][576,610]{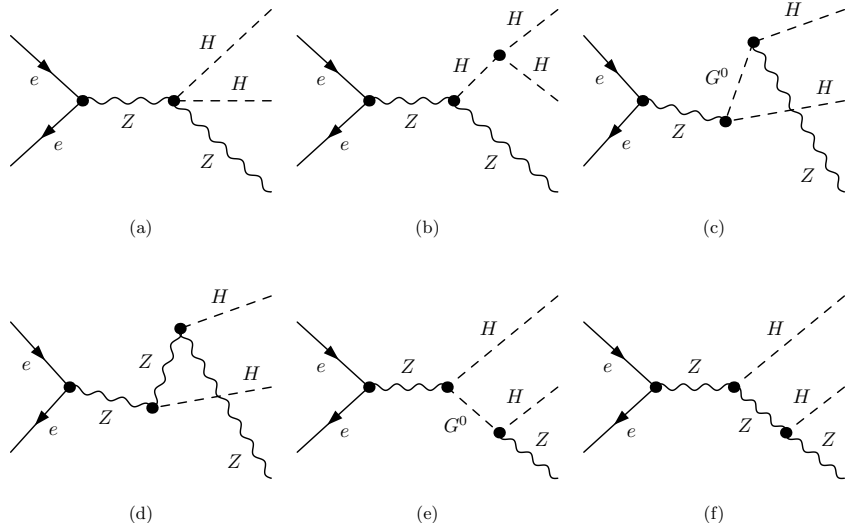}}
\caption{The tree level Feynman diagrams for $e^+ e^- \rightarrow
H H Z$.}
\end{figure}

\par
The ${\cal O}(\alpha_{ew})$ (one-loop level) virtual corrections to
$e^+ e^- \rightarrow H H Z$ can be expressed as
\begin{eqnarray}
\sigma_{{\rm virtual}} = \sigma_{{\rm tree}} \delta_{{\rm
virtual}} = \frac{(2 \pi)^4}{2 |\vec{p}_1| \sqrt{s}} \int {\rm d}
\Phi_3 \overline{\sum_{{\rm spin}}} {\rm Re} \left( {\cal M}_{{\rm
tree}} {\cal M}_{{\rm virtual}}^{\ast} \right),
\end{eqnarray}
where $\vec{p}_1$ is the c.m. momentum of the incoming positron,
${\rm d} \Phi_3$ is the three-body phase space element, and the
bar over summation recalls averaging over initial spins
\cite{particle data}. ${\cal M}_{{\rm virtual}}$ is the amplitude
of the virtual Feynman diagrams, including self-energy, vertex,
box, pentagon and counterterm diagrams. All the Feynman diagrams
and the relevant amplitudes are created by $FeynArts$ 3
\cite{Feynarts3}, and the Feynman amplitudes are subsequently
reduced by $FormCalc$ \cite{Formcalc}. Due to the fact that the
Yukawa coupling of Higgs/Goldstone to fermions is proportional to
the fermion mass, we neglect the contributions of the Feynman
diagrams which involve $H-e-\bar{e}$, $G^0-e-\bar{e}$,
$G^+-e-\bar{\nu}_{e}$ or $G^--\nu_{e}-\bar{e}$ vertex.

\par
As we know, the contributions of the virtual diagrams contain
both ultraviolet (UV) and infrared (IR) divergences, which can be
regularized by extending the dimensions of spinor and spacetime
manifolds to $D = 4 - 2 \epsilon$ \cite{tHooft and Veltman} and
giving the photon a fictitious mass $m_{\gamma}$, respectively. In
this paper, we adopt the complete on-mass-shell (COMS)
renormalization scheme \cite{COMS sheme} to fix all the
renormalization constants. All the tensor coefficients of the
one-loop integrals can be calculated by using the reduction
formulae presented in Refs. \cite{Passarino} and \cite{Dittmaier}.
As we expect, the UV divergence contributed by the loop diagrams
can be cancelled by that contributed by the counterterm diagrams
exactly, while the IR divergence still exists. Therefore, the
${\cal O}(\alpha_{ew}^4)$ virtual cross section $\sigma_{{\rm
virtual}}$ is independent of the UV regularization parameter
$C_{{\rm UV}} = 1/\epsilon - \gamma_E + {\rm log} 4 \pi$, but
still a function of the IR regularization parameter $m_{\gamma}$.

\par
In order to get an IR finite cross section for $e^+ e^-
\rightarrow H H Z$ up to the order of ${\cal O}(\alpha_{ew}^4)$,
we should consider the ${\cal O}(\alpha_{ew})$ corrections to $e^+
e^- \rightarrow H H Z$ due to real photon emission. By using the
general phase-space-slicing algorithm \cite{PSS}, the
contributions to the photon emission process $e^+ e^- \rightarrow
H H Z \gamma$ are divided into a soft and a hard
contribution,
\begin{eqnarray}
\sigma_{{\rm real}}
=
\sigma_{{\rm soft}} + \sigma_{{\rm hard}}
=
\sigma_{{\rm tree}}
\left(
\delta_{{\rm soft}} + \delta_{{\rm hard}}
\right),
\end{eqnarray}
where the ``soft'' and ``hard'' refer to the energy of the
radiated photon. The energy $E_{\gamma}$ of the
radiated photon in the c.m. frame is considered soft and hard if
$E_{\gamma} \leq \Delta E$ and $E_{\gamma} > \Delta E$,
respectively. Both $\sigma_{{\rm soft}}$ and $\sigma_{{\rm hard}}$
depend on the arbitrary soft cutoff $\Delta E/E_{b}$, where $E_b =
\sqrt{s}/2$ is the electron beam energy in the c.m. frame, but the
real cross section $\sigma_{{\rm real}}$ is cutoff independent. In
our calculations the soft cutoff $\Delta E/E_{b}$ is set to be
very small, therefore, the terms of order $\Delta E/E_{b}$ can be
neglected and the soft contribution can be evaluated by using the
soft photon approximation analytically \cite{COMS sheme}
\cite{soft r approximation}
\begin{eqnarray}
\label{soft part} {\rm d} \sigma_{{\rm soft}} = -{\rm d}
\sigma_{{\rm tree}}  \frac{\alpha_{ew}}{2 \pi^2}
\int_{|\vec{k}_{\gamma}| \leq \Delta E} \frac{{\rm d}^3 k_{\gamma}}{2
k^0_{\gamma}} \left( \frac{p_1}{p_1 \cdot k_{\gamma}} -
\frac{p_2}{p_2 \cdot k_{\gamma}} \right)^2.
\end{eqnarray}
Here $k^0_{\gamma} = \sqrt{|\vec{k}_{\gamma}|^2 + m_{\gamma}^2}$.
$k_{\gamma} = (k^0_{\gamma}, ~\vec{k}_{\gamma})$ is the four
momentum of the radiated photon, and $p_1$ and $p_2$ are the four
momenta of $e^+$ and $e^-$, respectively. As shown in Eq.
(\ref{soft part}), the soft contribution has a IR
singularity at $m_{\gamma} = 0$. The IR divergence from the soft
contribution cancels exactly that from the virtual corrections.
Therefore, $\sigma_{{\rm virtual} + {\rm soft}}$, the sum of the
${\cal O}(\alpha_{ew}^4)$ virtual and soft cross
sections, is independent of the infinitesimal photon mass
$m_{\gamma}$.

\par
The hard contribution is UV and IR finite. It can
be computed numerically by using standard Monte Carlo technique.
In this paper, our computation of the hard contribution
$\sigma_{{\rm hard}}$ is performed with the help of $CompHEP$
\cite{comphep}, which is a package for evaluation of tree level
Feynman diagrams and integration over multi-particle phase space
by using the adaptive multi-dimensional integration program
Vegas \cite{vegas}.

\par
Up to the order of ${\cal O}(\alpha_{ew}^4)$, the corrected cross
section for $e^+ e^- \rightarrow H H Z$ is just the sum of
the ${\cal O}(\alpha_{ew}^3)$ Born cross section
$\sigma_{{\rm tree}}$, the ${\cal O}(\alpha_{ew}^4)$ virtual cross
section $\sigma_{{\rm virtual}}$ and the ${\cal O}(\alpha_{ew}^4)$
real cross section $\sigma_{{\rm real}}$,
\begin{eqnarray}
\sigma_{{\rm tot}}
=
\sigma_{{\rm tree}} +
\sigma_{{\rm virtual}} +
\sigma_{{\rm real}}
=
\sigma_{{\rm tree}}
\left(
1 + \delta_{{\rm tot}}
\right),
\end{eqnarray}
where $\delta_{{\rm tot}} = \delta_{{\rm virtual}} + \delta_{{\rm
soft}} + \delta_{{\rm hard}}$ is the full ${\cal O}(\alpha_{ew})$
electroweak relative correction. As we expect, the corrected cross
section, $\sigma_{{\rm tot}}$, is independent of $C_{{\rm UV}}$
and $m_{\gamma}$, since it doesn't contains any UV or IR
singularity.

\par
To discuss the origin of some of the large corrections, we need to calculate
the photonic (QED) corrections and the genuine weak corrections separately.
The QED corrections comprise two parts:
the QED virtual corrections $\sigma_{{\rm virtual}}^{{\rm QED}}$
which contributed by the loop diagrams with virtual photon exchange in the
loop and the corresponding parts of the counterterms, and the real
corrections $\sigma_{{\rm real}}$.
Therefore the QED relative correction $\delta_{{\rm E}}$ can be expressed as
\begin{eqnarray}
\delta_{{\rm E}} = \delta_{{\rm virtual}}^{{\rm QED}} + \delta_{{\rm soft}} +
\delta_{{\rm hard}},
\end{eqnarray}
where $\delta_{{\rm virtual}}^{{\rm QED}}
= \sigma_{{\rm virtual}}^{{\rm QED}}/\sigma_{{\rm tree}}$, and the genuine weak
relative correction $\delta_{{\rm W}}$ is defined as
\begin{eqnarray}
\delta_{{\rm W}} = \delta_{{\rm tot}} - \delta_{{\rm E}}.
\end{eqnarray}

\section{Numerical results}
\label{sec-results}

For the numerical calculation we use the following SM input parameters \cite{particle data},
\begin{eqnarray}
m_e = 0.510998902~ {\rm MeV}, ~~~~~~
m_{\mu} = 105.658357~ {\rm MeV}, ~~~~~~
m_{\tau} = 1.77699~ {\rm GeV},~~~~~ \nb \\
m_u = 66~ {\rm MeV}, ~~~~~~~~~~~~~~~~~~~
m_c = 1.2~ {\rm GeV}, ~~~~~~~~~~~~~~~~~
m_t = 174.3~ {\rm GeV},~~~~~~~~ \nb \\
m_d = 66~ {\rm MeV}, ~~~~~~~~~~~~~~~~~~~
m_s = 150~ {\rm MeV}, ~~~~~~~~~~~~~~~~
m_b = 4.3~ {\rm GeV},~~~~~~~~~~~ \nb \\
m_W = 80.423~ {\rm GeV}, ~~~~~~~~~~~~~ m_Z = 91.1876~ {\rm GeV},
~~~~~~ \alpha_{ew}^{-1}(0) = 137.03599976. ~~~~~
\end{eqnarray}
By using the relevant SM parameters listed above, we obtain
\begin{eqnarray}
v = ( \sqrt{2} G_F )^{-1/2}
= m_W \sin\theta_W/\sqrt{\pi \alpha_{ew}}
= 250.356~ {\rm GeV}.
\end{eqnarray}
Besides these SM input parameters, six more input parameters
should be given in the numerical calculation, which are the c.m.
energy $\sqrt{s}$, the Higgs mass $m_{H}$, the mass parameter
of dimensional regularization $Q$,
the UV regularization parameter $C_{{\rm UV}}$,
the IR regularization parameter $m_{\gamma}$ and the
soft cutoff $\Delta E/E_{b}$. Since the corrected cross section
$\sigma_{{\rm tot}}$ is independent of the mass parameter of
dimensional regularization,
we set $Q = 1~ {\rm GeV}$ in the numerical calculation.

\par
In Table \ref{table-1} we present some numerical results of the
${\cal O}(\alpha_{ew}^4)$ cross section $\sigma_{{\rm virtual} +
{\rm soft}}$ for $e^+ e^- \rightarrow H H Z$, where the soft
cutoff $\Delta E/E_b$ and the UV regularization parameter $C_{{\rm
UV}}$ are set to be $\Delta E/E_b = 10^{-3}$ and $C_{{\rm UV}} =
0$, respectively. The middle two columns, labelled with I and II,
are corresponding to the cases of $m_{\gamma} = 10^{-20}~ {\rm
GeV}$ and $m_{\gamma} = 1~ {\rm GeV}$, respectively. Although the
Monte Carlo statistical error of the cross section $\sigma_{{\rm
virtual} + {\rm soft}}$ is the order of $10^{-3}$, we reserve the
output numbers in columns I and II with 14 digits. By comparing
the two columns of output numbers, we find that the results are
stable over 8 digits when varying the fictitious photon mass
$m_{\gamma}$ from $10^{-20}$ to $1~ {\rm GeV}$. Therefore, we draw
a conclusion that the ${\cal O}(\alpha_{ew}^4)$ cross section
$\sigma_{{\rm virtual} + {\rm soft}}$ is independent of
$m_{\gamma}$ within the statistical error. The results with 4 (or
3) significant digits and the corresponding Monte Carlo
integration errors are presented in the last column which is
labelled with RES(ERR).

\begin{table}
$$\begin{array}{c@{\quad}c@{\quad}c}
\hline
\sqrt{s}~[{\rm GeV}]
&
m_H~[{\rm GeV}]
&
\begin{array}{c}
\sigma_{{\rm virtual} + {\rm soft}}~[{\rm fb}] \\
~~~~~~~~~~~~~~
{\rm I}~~~~~~~~~~~~~~~~~~~~~~~~~~~~~~
{\rm II}~~~~~~~~~~~~~~~~~
{\rm RES}({\rm ERR})
\end{array} \\
\hline
500  & 115 & -0.13108006089013 ~~~~~ -0.13108006072333 ~~~~~ -0.1311(2) \\
     & 150 & -0.05089427858862 ~~~~~ -0.05089427861858 ~~~~~ -0.0509(1) \\
\hline
800  & 115 & -0.10810404820086 ~~~~~ -0.10810404820051 ~~~~~ -0.1081(3) \\
     & 150 & -0.08643256423041 ~~~~~ -0.08643256423012 ~~~~~ -0.0864(2) \\
\hline
1000 & 115 & -0.09123978783411 ~~~~~ -0.09123978977518 ~~~~~ -0.0912(3) \\
     & 150 & -0.08051191441546 ~~~~~ -0.08051191441530 ~~~~~ -0.0805(2) \\
\hline
2000 & 115 & -0.04961299068878 ~~~~~ -0.04961299068876 ~~~~~ -0.0496(2) \\
     & 150 & -0.04848771307528 ~~~~~ -0.04848771307526 ~~~~~ -0.0485(2) \\
\hline
\end{array}$$
\caption{The ${\cal O}(\alpha_{ew}^4)$ cross section $\sigma_{{\rm
virtual} + {\rm soft}}$ for $e^+ e^- \rightarrow H H Z$ process
for various Higgs boson mass (115 GeV and 150 GeV) and c.m. energy
values (500 GeV, 800 GeV, 1000 GeV and 2000 GeV).} \label{table-1}
\end{table}

\par
Analogously, the UV finiteness of $\sigma_{{\rm virtual}}$
can also be checked numerically. We find that the numerical results of
$\sigma_{{\rm virtual}}$ are stable over 7 digits when varying the UV
regularization parameter $C_{{\rm UV}}$ from 0 to $10^{5}$ for various
$\sqrt{s}$ and $m_H$. For simplicity, we do not present these numerical
results in this section.

\begin{figure}[htbp]
\centering
\scalebox{0.75}{\includegraphics*[70,71][550,412]{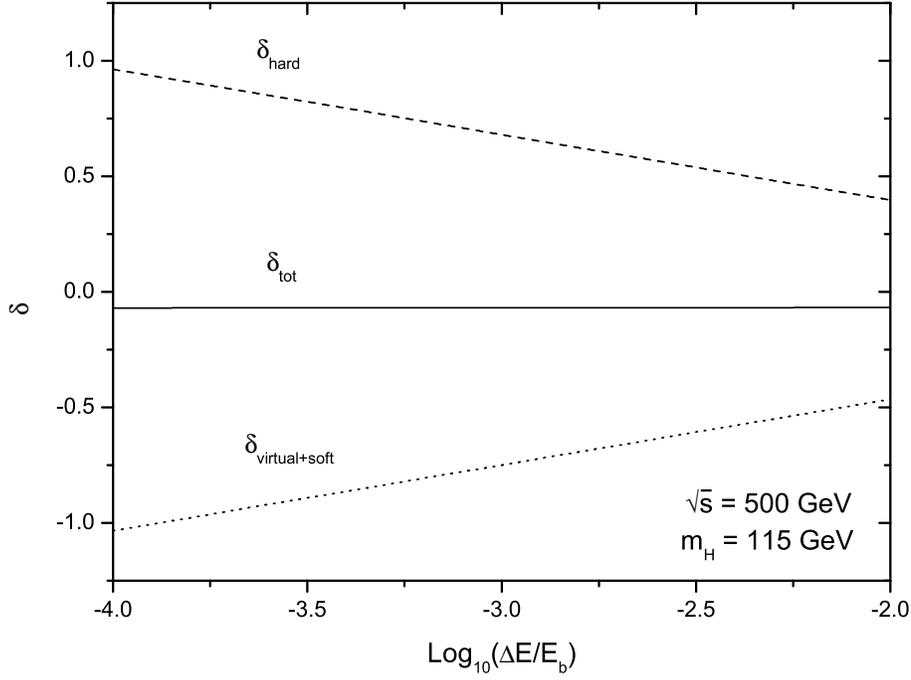}}
\caption{The ${\cal O}(\alpha_{ew})$ relative correction
to $e^+ e^- \rightarrow H H Z$ as a function of the soft
cutoff $\Delta E/E_b$.}
\end{figure}

\begin{figure}[htbp]
\centering
\scalebox{0.75}{\includegraphics*[70,71][550,412]{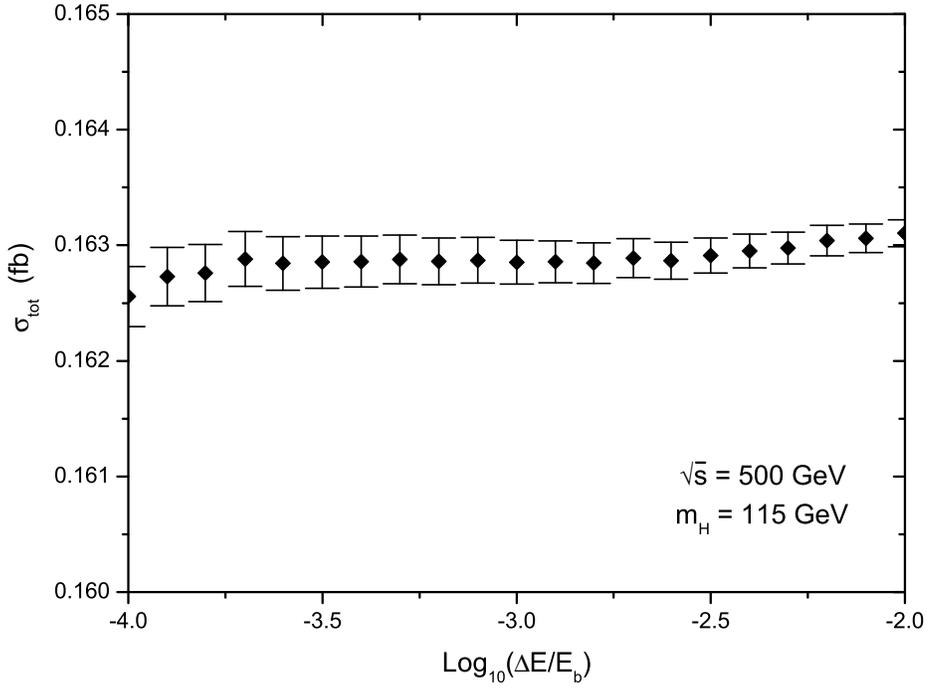}}
\caption{The dependence of the corrected cross section for
$e^+ e^- \rightarrow H H Z$ on the soft cutoff $\Delta E/E_b$.}
\end{figure}

\par
In Fig.2 we present the ${\cal O}(\alpha_{ew})$ relative
correction to $e^+ e^- \rightarrow H H Z$ as a function of the
soft cutoff $\Delta E/E_b$, assuming $m_{H} = 115~ {\rm GeV}$ and
$\sqrt{s} = 500~ {\rm GeV}$. As shown in this figure, both
$\delta_{{\rm virtual} + {\rm soft}}$ ($= \delta_{{\rm virtual}} +
\delta_{{\rm soft}}$) and $\delta_{{\rm hard}}$
depend on the soft cutoff $\Delta E/E_b$, but the full
${\cal O}(\alpha_{ew})$ electroweak relative correction $\delta_{{\rm
tot}}$ is cutoff independent. To show the cutoff independence more
clearly, we present $\sigma_{{\rm tot}}$, the corrected cross
section for $e^+ e^- \rightarrow H H Z$ which includes the full
${\cal O}(\alpha_{ew})$ electroweak corrections, with the
statistical errors from the Monte Carlo integration in Fig.3. As
shown in Fig.3, a clear plateau is reached for $\Delta E/E_b$ in
the range $10^{-4}-10^{-2}$ and the corrected cross section
$\sigma_{{\rm tot}}$ is obviously independent of $\Delta E/E_b$.

\par
Until now, we have checked the $m_{\gamma}$, $C_{{\rm UV}}$ and
$\Delta E/E_b$ independence of the full ${\cal O}(\alpha_{ew})$
electroweak corrections to $e^+ e^- \rightarrow H H Z$. In the
following calculation, $m_{\gamma}$, $C_{{\rm UV}}$ and $\Delta E/E_b$
are fixed to be $10^{-2}~ {\rm GeV}$, 0 and $10^{-3}$, respectively.

\begin{figure}[htbp]
\centering
\scalebox{0.75}{\includegraphics*[70,71][550,412]{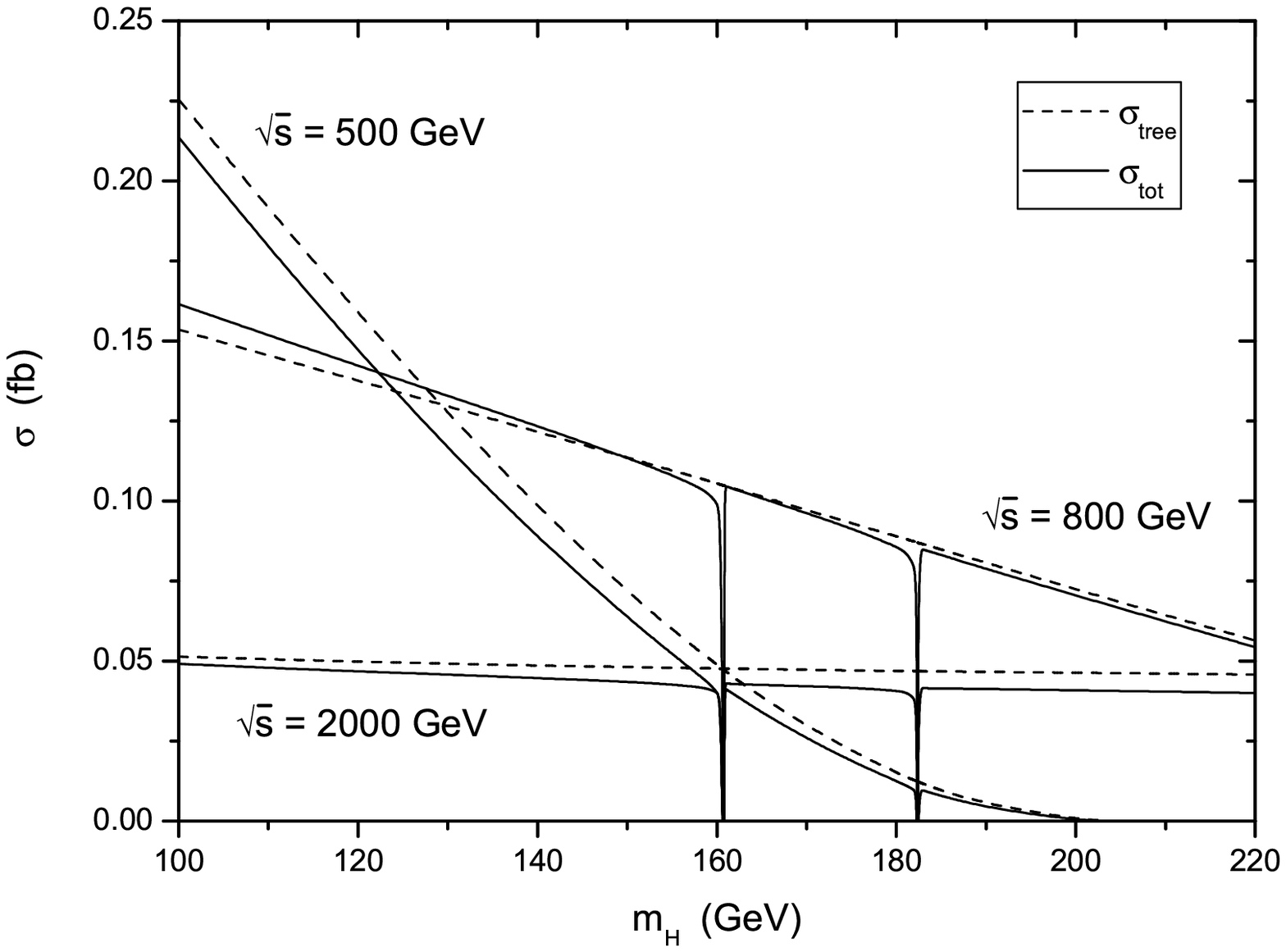}}
\caption{The Born cross section and the corrected cross
section for $e^+ e^- \rightarrow H H Z$ as functions of the
Higgs boson mass.}
\end{figure}

\begin{figure}[htbp]
\centering
\scalebox{0.75}{\includegraphics*[70,71][550,412]{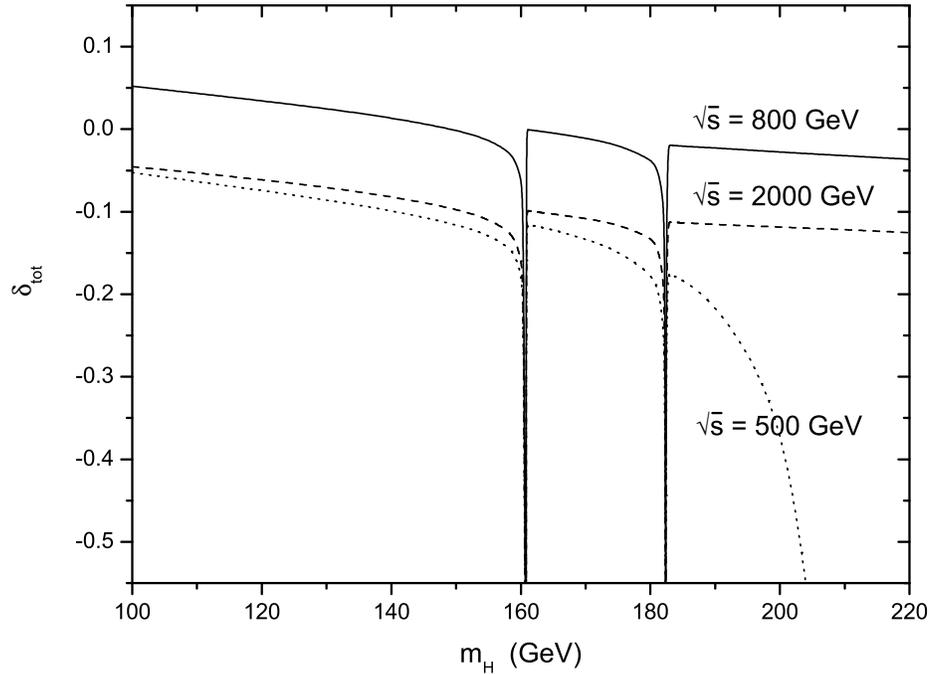}}
\caption{The full ${\cal O}(\alpha_{ew})$ electroweak relative correction to
$e^+ e^- \rightarrow H H Z$ as a function of the Higgs boson mass.}
\end{figure}

\par
In Fig.4 we present the Born cross section $\sigma_{{\rm tree}}$
and the corrected cross section $\sigma_{{\rm tot}}$ for $e^+ e^-
\rightarrow H H Z$ as functions of the Higgs boson mass $m_H$
for $\sqrt{s} = 500$, $800$ and $2000~{\rm GeV}$, respectively. As
shown in this figure, each solid curve has two spikes at the
vicinities of $m_H = 2 m_W$ and $m_H = 2 m_Z$, which just reflect
the resonance effects at $m_H = 2 m_W$ and $m_H = 2 m_Z$, respectively.
For $\sqrt{s} = 2000~ {\rm GeV}$, both
$\sigma_{{\rm tree}}$ and $\sigma_{{\rm tot}}$ are insensitive to
$m_H$, decrease very slowly as the increment of
$m_H$ from $100$ to $220~{\rm GeV}$. In contrast to the
case of $\sqrt{s} = 2000~ {\rm GeV}$, the cross sections are
sensitive to $m_H$ when $\sqrt{s} = 500~ {\rm GeV}$. They decrease
rapidly to zero as $m_H$ increases to about $204~{\rm GeV}$.

\par
To describe the full ${\cal O}(\alpha_{ew})$ electroweak
corrections to the Born cross section for $e^+ e^- \rightarrow H H
Z$ quantitatively, we plot the full ${\cal O}(\alpha_{ew})$
relative correction $\delta_{{\rm tot}}$, defined as $\delta_{{\rm
tot}} = (\sigma_{{\rm tot}} - \sigma_{{\rm tree}})/\sigma_{{\rm
tree}}$, as a function of $m_H$ in Fig.5. For $\sqrt{s} = 500~
{\rm GeV}$, which is the most favorable c.m. energy for $e^+ e^-
\rightarrow H H Z$ with intermediate Higgs boson mass, the
relative correction is negative in the range
of $100~ {\rm GeV} < m_H < 204~ {\rm GeV}$.
It decreases from $-5.3\%$ to $-11.50\%$ as $m_H$ increases
from 100 to 150 GeV.
Since the cross section near threshold is very small, the large
relative correction in this region is phenomenologically insignificant.
For $\sqrt{s} = 2000~ {\rm GeV}$, the relative
correction is also negative in the range of $100~ {\rm GeV} < m_H
< 220~ {\rm GeV}$. It decreases from $-4.6\%$ to $-12.5\%$ as the
increment of $m_H$ from $100$ to $220{\rm GeV}$. For $\sqrt{s} =
800~ {\rm GeV}$, the relative correction is positive when $m_H <
150~ {\rm GeV}$. It varies from $5.2\%$ to $-3.7\%$ as $m_H$ running
from 100 to 220 GeV. The numerical results of $\sigma_{{\rm
tree}}$, $\sigma_{{\rm tot}}$ and $\delta_{{\rm tot}}$ for some
typical values of $m_H$ and $\sqrt{s}$ are presented in Table
\ref{table-2}.

\begin{table}
\newcommand{\phm}{\phantom{-}}
$$\begin{array}{c@{\quad}c@{\quad}l@{\quad}l@{\quad}l}
\hline
\sqrt{s}~ [{\rm GeV}]
&
m_H~ [{\rm GeV}] &
\sigma_{{\rm tree}}~ [{\rm fb}]
&
\sigma_{{\rm tot}}~ [{\rm fb}]
&
\delta_{{\rm tot}}~ [\%] \\
\hline
    & 115 & 0.17493(2) & 0.1629(2) & -6.9(1) \\
500 & 150 & 0.071834(6) & 0.06357(6) & -11.50(7) \\
    & 200 & 0.49611(3) \cdot 10^{-3} & 0.3329(2) \cdot 10^{-3} & -32.90(4) \\
\hline
    & 115 & 0.17428(2) & 0.1740(3) & -0.2(1) \\
600 & 150 & 0.10840(1) & 0.1041(1) & -4.0(1) \\
    & 200 & 0.031802(3) & 0.02935(2) & -7.71(7) \\
\hline
    & 115 & 0.15868(3) & 0.1632(3) & \phm2.8(2) \\
700 & 150 & 0.11665(2) & 0.1155(2) & -1.0(1) \\
    & 200 & 0.058846(7) & 0.05665(7) & -3.7(1) \\
\hline
    & 115 & 0.14156(3) & 0.1471(3) & \phm3.9(2) \\
800 & 150 & 0.11363(2) & 0.1135(2) & -0.1(2) \\
    & 200 & 0.07246(1) & 0.0705(1) & -2.7(1) \\
\hline
     & 115 & 0.11293(2) & 0.1168(3) & \phm3.4(3) \\
1000 & 150 & 0.09890(2) & 0.0983(3) & -0.6(2) \\
     & 200 & 0.07790(1) & 0.0753(2) & -3.3(2) \\
\hline
     & 115 & 0.07119(2) & 0.0704(3) & -1.1(4) \\
1500 & 150 & 0.06684(2) & 0.0634(2) & -5.1(3) \\
     & 200 & 0.06165(1) & 0.0569(2) & -7.7(3) \\
\hline
     & 115 & 0.05021(1) & 0.0473(2) & -5.8(4) \\
2000 & 150 & 0.04812(1) & 0.0435(2) & -9.6(4) \\
     & 200 & 0.04630(1) & 0.0408(2) & -11.9(4) \\
\hline
\end{array}$$
\caption{The Born cross section $\sigma_{{\rm tree}}$, the
corrected cross section $\sigma_{{\rm tot}}$ and the full ${\cal
O}(\alpha_{ew})$ electroweak relative correction $\delta_{{\rm
tot}}$ for various Higgs boson mass and c.m. energy values.}
\label{table-2}
\end{table}

\begin{figure}[htbp]
\centering
\scalebox{0.75}{\includegraphics*[70,71][550,412]{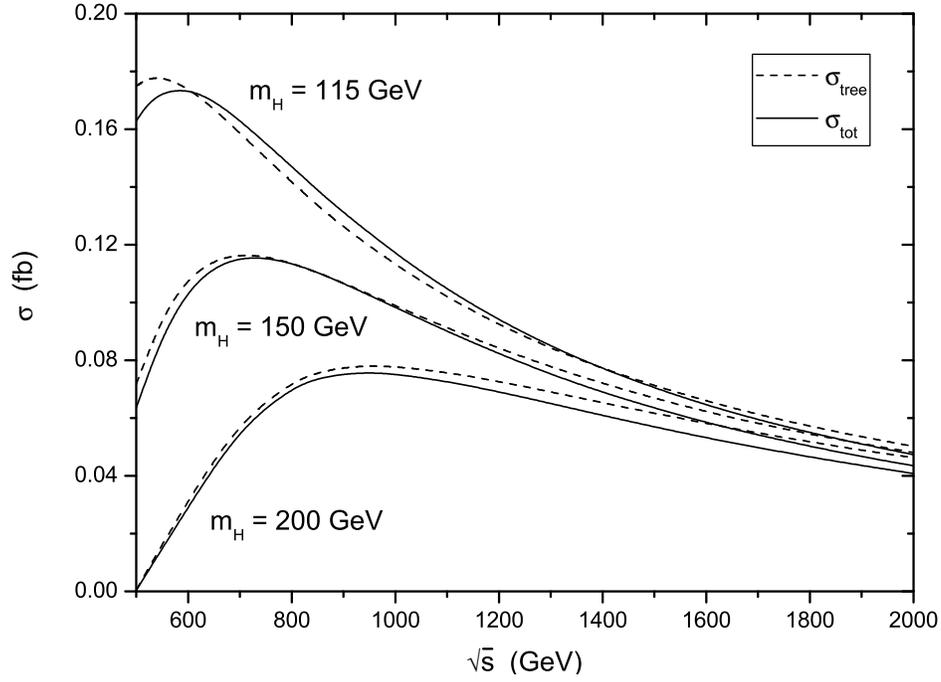}}
\caption{The Born cross section and the corrected cross
section for $e^+ e^- \rightarrow H H Z$ as functions of the
c.m. energy.}
\end{figure}

\begin{figure}[htbp]
\centering
\scalebox{0.75}{\includegraphics*[70,71][550,412]{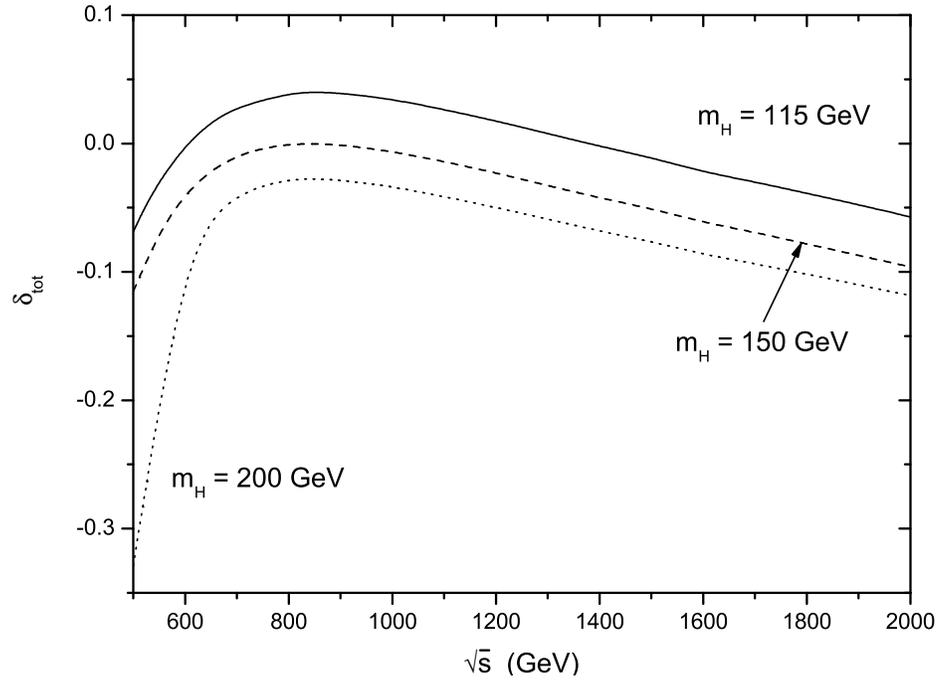}}
\caption{The full ${\cal O}(\alpha_{ew})$ electroweak relative correction to
$e^+ e^- \rightarrow H H Z$ as a function of the c.m. energy.}
\end{figure}

\par
In Fig.6 we depict the Born cross section $\sigma_{{\rm tree}}$
and the corrected cross section $\sigma_{{\rm tot}}$ for $e^+ e^-
\rightarrow H H Z$ as functions of the c.m. energy $\sqrt{s}$ for
$m_H = 115$, 150 and 200 GeV, respectively. The c.m. energy varies
in the range $500~ {\rm GeV} < \sqrt{s} < 2000~ {\rm GeV}$, which
is accessible at future linear colliders, such as TESLA
\cite{tesla}, NLC \cite{nlc}, JLC \cite{jlc} and CERN CLIC
\cite{clic}. From this figure we find that both the Born cross
section $\sigma_{{\rm tree}}$ and the corrected cross section
$\sigma_{{\rm tot}}$ increase firstly, reach their maximal values,
and then decrease with the increment of $\sqrt{s}$. For $m_H =
115~{\rm GeV}$, $\sigma_{{\rm tree}}$ and $\sigma_{{\rm tot}}$
reach their maximal values of about 0.178 and 0.174 fb at
$\sqrt{s} \sim 550~ {\rm GeV}$ and $\sqrt{s} \sim 600~ {\rm GeV}$,
respectively. For $m_H = 150~ {\rm GeV}$ $\sigma_{{\rm tree}}$ and
$\sigma_{{\rm tot}}$ reach the maximums of about 0.117 and 0.116
fb at $\sqrt{s} \sim 700~ {\rm GeV}$, and for $m_H = 200~ {\rm
GeV}$ they reach about 0.078 and 0.076 fb at $\sqrt{s} \sim 950~
{\rm GeV}$, respectively.

\par
The dependence of the full ${\cal O}(\alpha_{ew})$ electroweak
relative correction to $e^+ e^- \rightarrow H H Z$ on the c.m.
energy $\sqrt{s}$ is displayed in Fig.7. As shown in this figure,
the full ${\cal O}(\alpha_{ew})$ electroweak corrections suppress
the Born cross section in the c.m. energy range of $500~ {\rm GeV}
< \sqrt{s} < 2000~ {\rm GeV}$ for $m_H = 150$ and 200 GeV, while
enhance the Born cross section in the c.m. energy range of $610~
{\rm GeV} < \sqrt{s} < 1360~ {\rm GeV}$ for $m_H = 115~ {\rm
GeV}$. The relative corrections can reach about $-6.9\%$ and
$-11.5\%$ at $\sqrt{s} = 500~ {\rm GeV}$ for $m_H = 115$ and 150
GeV respectively. For $m_H = 200~ {\rm GeV}$, the relative
correction is large and can exceed $-10\%$. It ranges from
$-2.7\%$ to $-11.9\%$ as $\sqrt{s}$ varying in the range of $800~
{\rm GeV} < \sqrt{s} < 2000~ {\rm GeV}$. We can see that in some
parameter space the electroweak relative corrections are only few
percent and might be below the achievable experimental accuracy.

\begin{figure}[htbp]
\centering
\scalebox{0.81}{\includegraphics*[37,37][556,350]{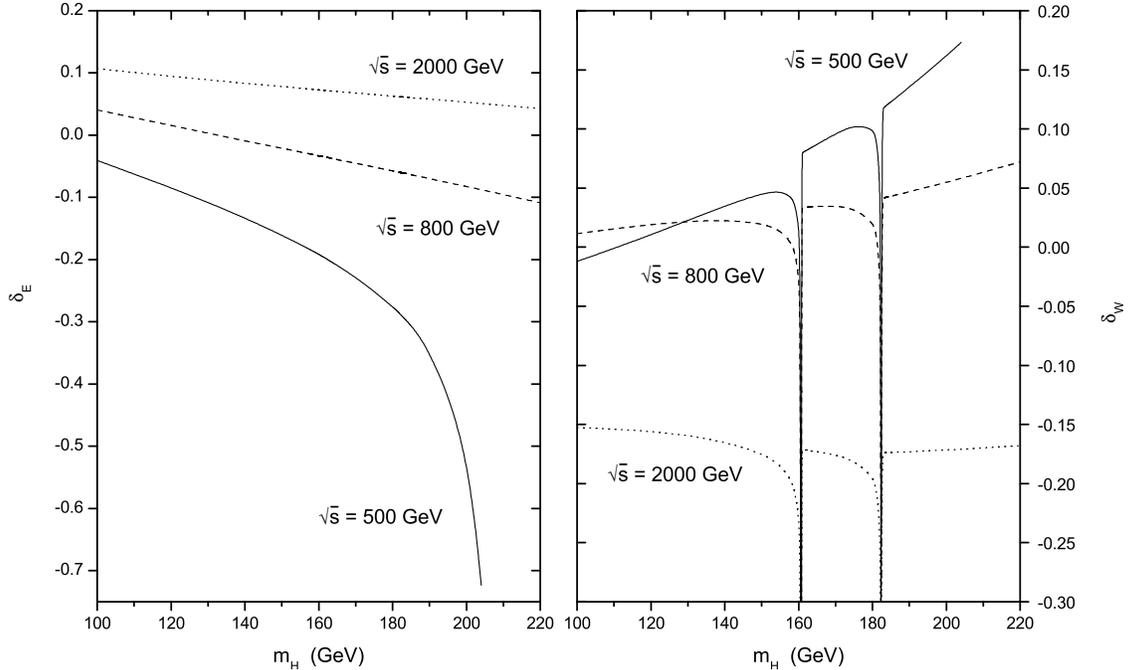}}
\caption{The dependence of the QED relative correction $\delta_E$
and the genuine weak relative correction $\delta_W$ to $e^+ e^-
\rightarrow H H Z$ on the Higgs boson mass.}
\end{figure}

\par
In Fig.8 we present the dependence of the QED relative correction
$\delta_{\rm E}$ and the genuine weak relative correction $\delta_{\rm W}$ to
$e^+ e^- \rightarrow H H Z$ on the Higgs boson mass separately.
From this figure we find that at threshold the genuine weak
relative correction is a striking contrast to the QED relative
correction, and the contribution of the ${\cal O}(\alpha_{ew})$
electroweak correction is overwhelmingly dominanted by the QED
correction. For $\sqrt{s} = 500~ {\rm GeV}$, the genuine weak
relative correction strongly depends on the Higgs boson mass. It
increases from $-1.20\%$ to $4.56\%$ as $m_H$ increases from 100
to 150 GeV. For $\sqrt{s} = 2000$ GeV, the genuine weak relative
correction is insensitive to the Higgs boson mass. It is about
$-15\% \sim -17\%$ for $m_H$ in the range of $100~ {\rm GeV} < m_H
< 220~ {\rm GeV}$.

\par
The $\sqrt{s}$ dependence of the QED relative correction
$\delta_{\rm E}$ and genuine weak relative correction $\delta_{\rm W}$ to $e^+
e^- \rightarrow H H Z$ are displayed in Fig.9. Together with Fig.6
we can see from this figure that for the range of $\sqrt{s}$ where
the cross section is relatively large, the QED relative correction
is not too large and the genuine weak relative correction is less
than $5\%$. For $\sqrt{s}
> 1000~ {\rm GeV}$, the Higgs mass dependence of the genuine weak
relative correction is small. As $\sqrt{s}$ increases to $2000~
{\rm GeV}$, the genuine weak relative correction can reach about
$-15\% \sim -18\%$ for $m_H$ in the range of $115 ~ {\rm GeV} <
m_H < 200 ~ {\rm GeV}$.

\begin{figure}[htbp]
\centering
\scalebox{0.81}{\includegraphics*[37,37][556,350]{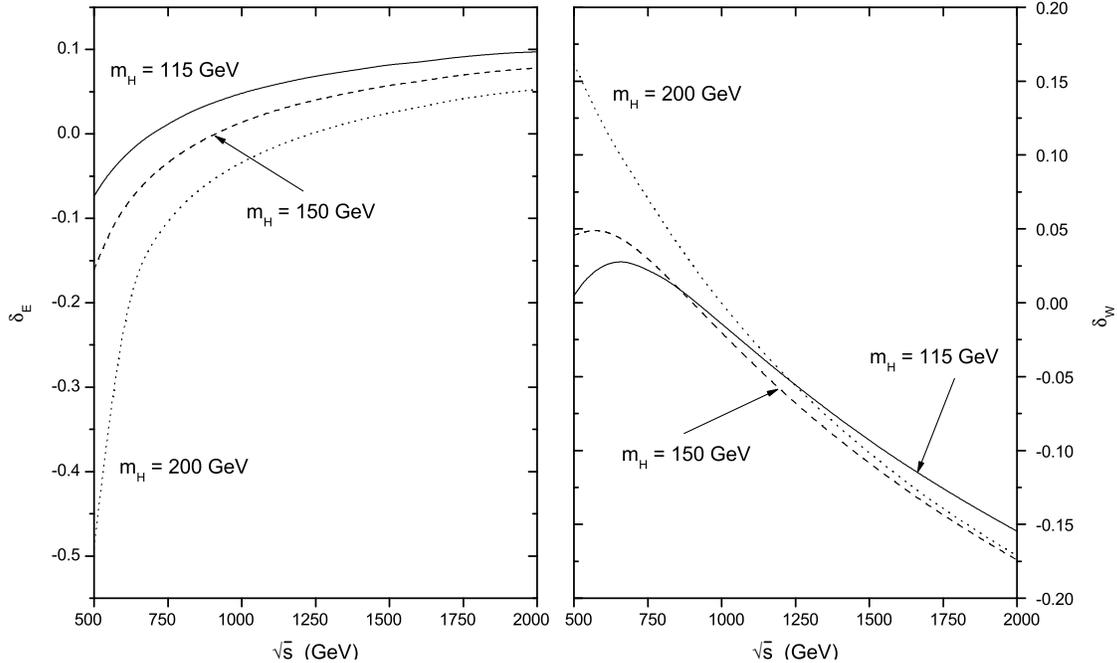}}
\caption{The dependence of the QED relative correction and the genuine weak relative
correction to $e^+ e^- \rightarrow H H Z$ on the c.m. energy.}
\end{figure}

\section{ Summary}
\label{sec-summary}

In this paper we calculate the full ${\cal O}(\alpha_{ew})$
electroweak corrections to $e^+ e^- \rightarrow H H Z$ at a LC in
the SM, and analyze the dependence of the Born cross section, the
corrected cross section including full ${\cal O}(\alpha_{ew})$
electroweak corrections and the relative correction on the $m_H$
and $\sqrt{s}$. To see the origin of some of the large corrections
clearly, we calculate the QED and genuine weak corrections
separately. From the numerical results we find that the full
${\cal O}(\alpha_{ew})$ electroweak corrections significantly
suppress or enhance the Born cross section, depending on the Higgs
boson mass and the c.m. energy of a LC. Both the Born cross
section and the corrected cross section are insensitive to the
Higgs boson mass in the range of $100~ {\rm GeV} < m_H < 220~ {\rm
GeV}$ for $\sqrt{s} = 2000~ {\rm GeV}$, but strongly related to
the Higgs boson mass in the range of $100~ {\rm GeV} < m_H < 204~
{\rm GeV}$ for $\sqrt{s} = 500~ {\rm GeV}$. With our chosen
$\sqrt{s}-m_H$ parameter space in this paper, the relative
corrections are a few percent generally, and can exceed $-10\%$
when $\sqrt{s} \sim 500~ {\rm GeV}$ and $m_H \sim 150~ {\rm GeV}$.
Therefore, the full ${\cal O}(\alpha_{ew})$ electroweak
corrections should be taken into account in the precise experiment
analysis. We should also mention that in some parameter space,
where the cross section is sizeable, the total relative
corrections are only few percent and thus probably might be below
the achievable experimental accuracy.

\vskip 10mm
\noindent{\large\bf Note Added:} As we were amending
this manuscript, we became aware of a similar paper by G.
Belanger, et al.,\cite{Belanger}. They presented a numerical
comparison in Table 3 of Ref.\cite {Belanger}.

\vskip 10mm
\noindent{\large\bf Acknowledgments:} This work was
supported in part by the National Natural Science Foundation of
China and  a grant from the University of Science and Technology
of China.

\vskip 10mm
\begin{flushleft} {\bf Figure Captions} \end{flushleft}
\par
{\bf Figure 1} The tree level Feynman diagrams for $e^+ e^- \rightarrow H H Z$.
\par
{\bf Figure 2} The ${\cal O}(\alpha_{ew})$ relative correction
to $e^+ e^- \rightarrow H H Z$ as a function of the soft cutoff $\Delta E/E_b$.
\par
{\bf Figure 3} The dependence of the corrected cross section for
$e^+ e^- \rightarrow H H Z$ on the soft cutoff $\Delta E/E_b$.
\par
{\bf Figure 4} The Born cross section and the corrected cross
section for $e^+ e^- \rightarrow H H Z$ as functions of the
Higgs boson mass.
\par
{\bf Figure 5} The full ${\cal O}(\alpha_{ew})$ electroweak relative correction to
$e^+ e^- \rightarrow H H Z$ as a function of the Higgs boson mass.
\par
{\bf Figure 6} The Born cross section and the corrected cross
section for $e^+ e^- \rightarrow H H Z$ as functions of the
c.m. energy.
\par
{\bf Figure 7} The full ${\cal O}(\alpha_{ew})$ electroweak relative correction to
$e^+ e^- \rightarrow H H Z$ as a function of the c.m. energy.
\par
{\bf Figure 8} The dependence of the QED relative correction and the genuine weak
relative correction to $e^+ e^- \rightarrow H H Z$ on the Higgs boson mass.
\par
{\bf Figure 9} The dependence of the QED relative correction and the genuine weak
relative correction to $e^+ e^- \rightarrow H H Z$ on the c.m. energy.
\par
{\bf Table 1} The ${\cal O}(\alpha_{ew}^4)$ cross section
$\sigma_{{\rm virtual+soft}}$ for $e^+ e^- \rightarrow H H Z$
process for various Higgs boson mass (115 GeV and 150 GeV) and
c.m. energy values (500 GeV, 800 GeV, 1000 GeV and 2000 GeV).
\par
{\bf Table 2} The Born cross section $\sigma_{{\rm tree}}$, the
corrected cross section $\sigma_{{\rm tot}}$ and the full ${\cal
O}(\alpha_{ew})$ electroweak relative correction $\delta_{{\rm
tot}}$ for various Higgs boson mass and c.m. energy values.
\end{document}